\documentclass[5p]{elsarticle} 

\usepackage{epstopdf}
\usepackage{soul}
\usepackage[dvipsnames,usenames]{color}

\usepackage{microtype}
\usepackage{subcaption} 

\newcommand{\rub}{Ru$_7$B$_3$}
\newcommand{\rube}{Ru$_7\,^{11}$B$_3$}
\newcommand{\eb}{$^{11}$B}

\usepackage{graphicx}

\usepackage{amssymb}

\journal{Journal of Crystal Growth}

\begin{document}

\begin{frontmatter}



\title{Crystal growth and properties of the non-centrosymmetric superconductor, {\rub}}


\author{R.P. Singh\fnref{fn1}} 
\author{N.A. Parzyk}
\author{M.R. Lees}
\author{D.M$^\textrm{c}$K Paul}
\author{G. Balakrishnan\corref{cor1}}
\ead{G.Balakrishnan@warwick.ac.uk}

\cortext[cor1]{Corresponding author}
\fntext[fn1]{Present address: Department of Physics, IISER Bhopal, MP-462023, India}

\address{ Department of Physics, University of Warwick, Coventry CV4 7AL, UK}

\begin{abstract}

We describe the crystal growth of high quality single crystals of the  non-centrosymmetric superconductor, {\rub} by the floating zone technique, using an optical furnace equipped with xenon arc lamps. The crystals obtained are large and suitable for detailed measurements, and have been examined using x-ray Laue patterns. The superconducting properties of the crystals obtained have been investigated by magnetisation and resistivity measurements. Crystals have also been grown starting with enriched {\eb} isotope, making them suitable for neutron scattering experiments.

\end{abstract}


\end{frontmatter}



\section{Introduction}

In recent years, experimental and theoretical studies of the   non-\-centrosy\-mme\-tric superconductors (NCS) have attracted much attention due to their complex superconducting properties~\cite{Sigrist,BSbook}. Following the first report of the observation of superconductivity in the non-centrosymmetric heavy fermion compound CePt$_3$Si~\cite{Bauer}, there has been a lot of interest in NCS. In NCS, the crystal structure lacks a centre of inversion,  which implies that in these superconductors parity is no longer a conserved quantum number and a mixed singlet-triplet superconducting wave function is possible~\cite{Hillier,Luke}. As a result, NCS superconductors show significantly different properties from conventional superconducting systems such as large Pauli-limiting fields and helical vortex states.  Theoretical predictions  also suggest that NCS  can be candidates for topological materials~\cite{Sato} due to their strong asymmetric spin-orbit interaction. Despite the theoretical predictions, there are very few experimental studies reported on single crystals of NCS due to the lack of high quality single crystals of several of these materials.

{\rub} forms in a non-centrosymmetric hexagonal \linebreak
 {Th$_7$Fe$_3$}-type crystal structure with the space group \textit{P6$_3$mc} (No.~186) \cite{AronssonStructure,Matthias}. The superconducting transition temperatures reported for {\rub} in the literature vary from $\sim$2.5 to $\sim$3.4~K. For polycrystalline samples, reported values of superconducting transition temperature $T_c$ range from  {$\sim$}2.58~to~3.38~K~\cite{Fang}.  Previous attempts to obtain single crystals of this material have been successful using the Czochralski method in a tetra-arc furnace~\cite{Kase}. Kase~\textit{et al}.  report the superconducting transition temperature for their crystals to be different to their polycrystalline starting material~\cite{Kase}. The Ru-B binary phase diagram suggests that the {\rub} phase melts congruently at {$\sim$}1600~$^{\circ}$C~\cite{phasediag}.  In this paper, we report the growth of large single crystals of the non-centrosymmetric superconducting compound, {\rub}~by the floating zone technique using an optical mirror furnace~\cite{SinghNbRh} and present magnetisation and resistivity measurements made on these crystals to determine their superconducting properties.  We have used a four mirror optical furnace equipped with high power xenon arc lamps (4~x~3~kW) to grow these single crystals. This furnace has the capability of melting materials with melting points of up to a maximum of 2800~$^{\circ}$C. The crystals obtained by this technique are large enough for most physical property measurements and are ideal for neutron scattering experiments, where large volumes of single crystal are essential. We have also produced single crystals starting with the isotopically enriched {\eb}, which has a much reduced absorption cross section for neutrons in comparison to the normally abundant boron. These crystals can be used to investigate the vortex state of this superconductor using  neutron scattering techniques.

\section{Experimental Procedure}
 
Polycrystalline samples of {\rub} were prepared by arc melting stoichiometric quantities of high purity powders of Ru~(99.99\%), and either B~(99.9\%) or enriched boron isotope {\eb}~(99.52~At\%), on a water cooled copper hearth under a high purity argon gas (5N) atmosphere in a tri arc furnace (Centorr,~USA). To ensure phase homogeneity, the resulting buttons were remelted and flipped several times. The observed weight loss during the melting was negligible. The phase purity of the polycrystalline buttons was checked by powder x-ray diffraction. Several polycrystalline buttons obtained in this manner were used to cast rods ({$\sim$}40~mm in length and {$\sim$}6~mm in diameter) under flowing argon gas in a tri-arc furnace. Two different types of feed rods were made, one set of rods starting with B and another set made starting with {\eb}.  These rods were used as feed rods for the crystal growth in a high temperature optical furnace equipped with four 3~kW xenon arc lamps (CSI Model FZT-1200-X-VI~VP). The first crystal growth was carried out using a tungsten rod as the seed rod and a crystal obtained from the initial growths  was used as the seed for subsequent growths. Prior to starting the crystal growth, the growth chamber was flushed several times with high purity argon gas and then evacuated to a vacuum of  {$\sim$}10$^{-6}$~mbar and finally filled with argon gas to a pressure of 0.3~MPa for the growth. The crystal growth was carried out under a flow (1~l/min) of high purity argon gas. The feed and seed rods were counter rotated at 25--30~rpm and crystal growths were carried out at speeds of 3~to~6~mm/h. Crystals of {\rube} with the isotopically enriched boron, were grown under similar growth conditions.

\begin{figure}[!t]
\begin{centering}
\includegraphics[width=\columnwidth]{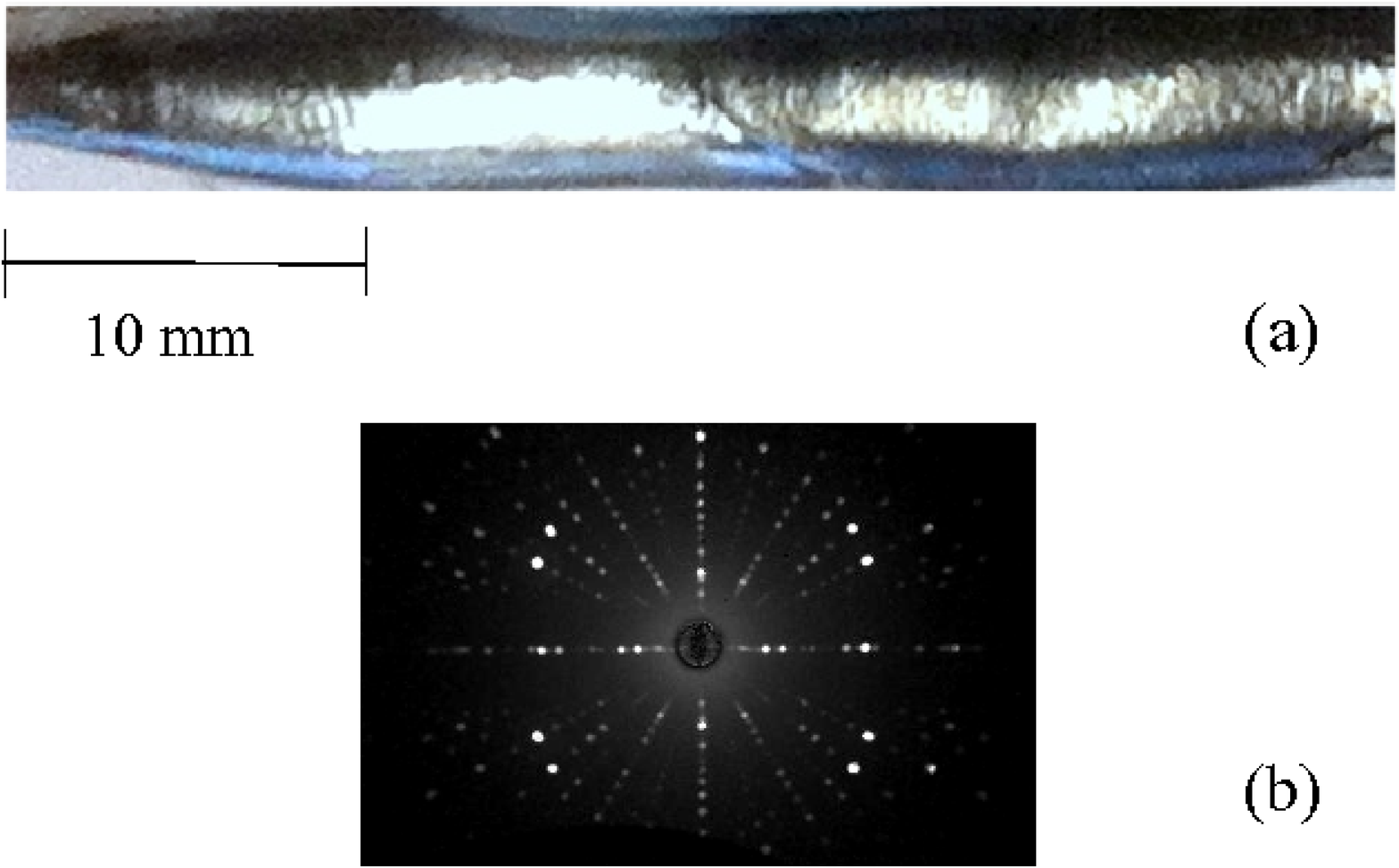}%
\caption{(a) Single crystal of {\rub}. (b)~Laue x-ray diffraction pattern of a piece cut out of the as grown boule along the [100] direction.}
\label{fig1}%
\end{centering}
\end{figure}

The crystal boules produced were first examined using Laue x-ray (Cu)  diffraction to check the quality of the crystals. Spark erosion was used to cut slices from the grown boule for subsequent measurements. Temperature and field dependent magnetisation measurements were made in the temperature range 1.8~to~300~K, using a Quantum Design MPMS-5S SQUID magnetometer. AC resistivity measurements in the temperature range 1.8 to 300~K were carried out on rectangular bar shaped samples ({$\sim$}4~mm in length and 
2~x~1~mm$^2$ in cross section) by the standard four-probe technique using a Quantum Design Physical Property Measurement System (PPMS). 

\section{Results and Discussion}
{\rub} melts congruently and therefore lends itself to crystal growth by the floating zone method. The crystals obtained starting with both the {\eb} as well as the $^{10}$B were identical in appearance and both the boules had shiny surfaces with a golden lustre. Crystals about 40 to 50~mm long and roughly 4~to~6~mm in diameter were obtained. As the melting temperature for this material is around 1600~$^{\circ}$C, the power of the xenon lamps required to maintain a stable molten zone was in the region of 15~to~20\%. 
 We found that in order to obtain the best crystals, slow growth rates of around {$\sim$}3~mm/h were necessary. There was evidence of boron evaporation during the growth process in all the growths conducted. The crystals were cut using a spark cutter. Prior to cutting, the crystals were examined using Laue x-ray back reflection. Laue x-ray photographs were taken along the length of each crystal on several faces to confirm the crystal quality. A photograph of the as grown {\rub} crystal is shown in Figure~\ref{fig1}(a). The Laue x-ray pattern obtained on a piece of the crystal cut out of the boule for measurements along the [100] direction is shown in Figure~\ref{fig1}(b). By use of the Laue x-ray diffraction, crystals oriented along particular crystallographic axes were cut from the as-grown boules for measurements.

     \begin{figure}[!t]
    \centering
    \begin{subfigure}[h]{0.7\columnwidth}
        \centering
        \includegraphics[width=\textwidth]{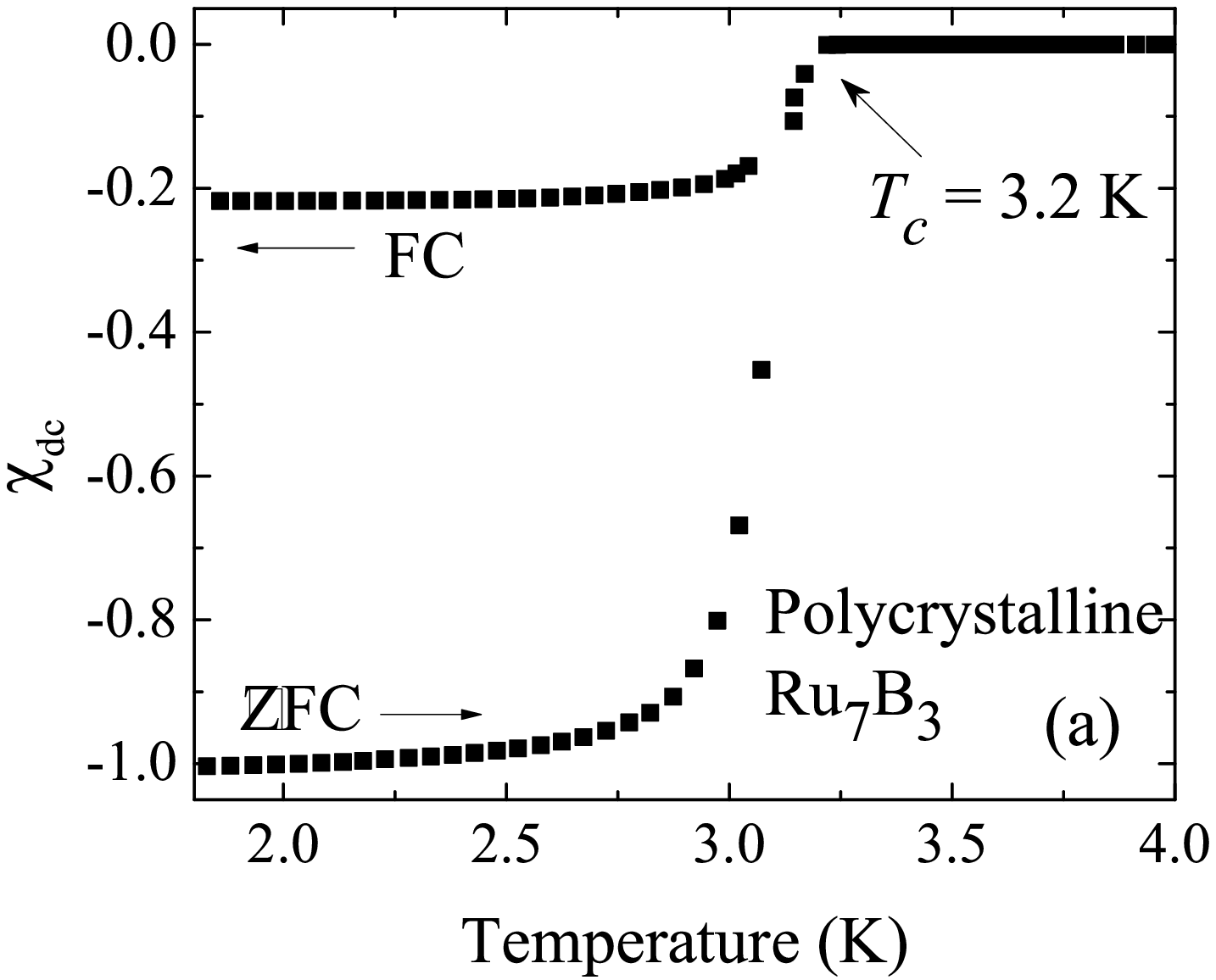}
        \label{fig2a}
    \end{subfigure}
    \begin{subfigure}[h]{0.7\columnwidth}
        \centering
        \includegraphics[width=\textwidth]{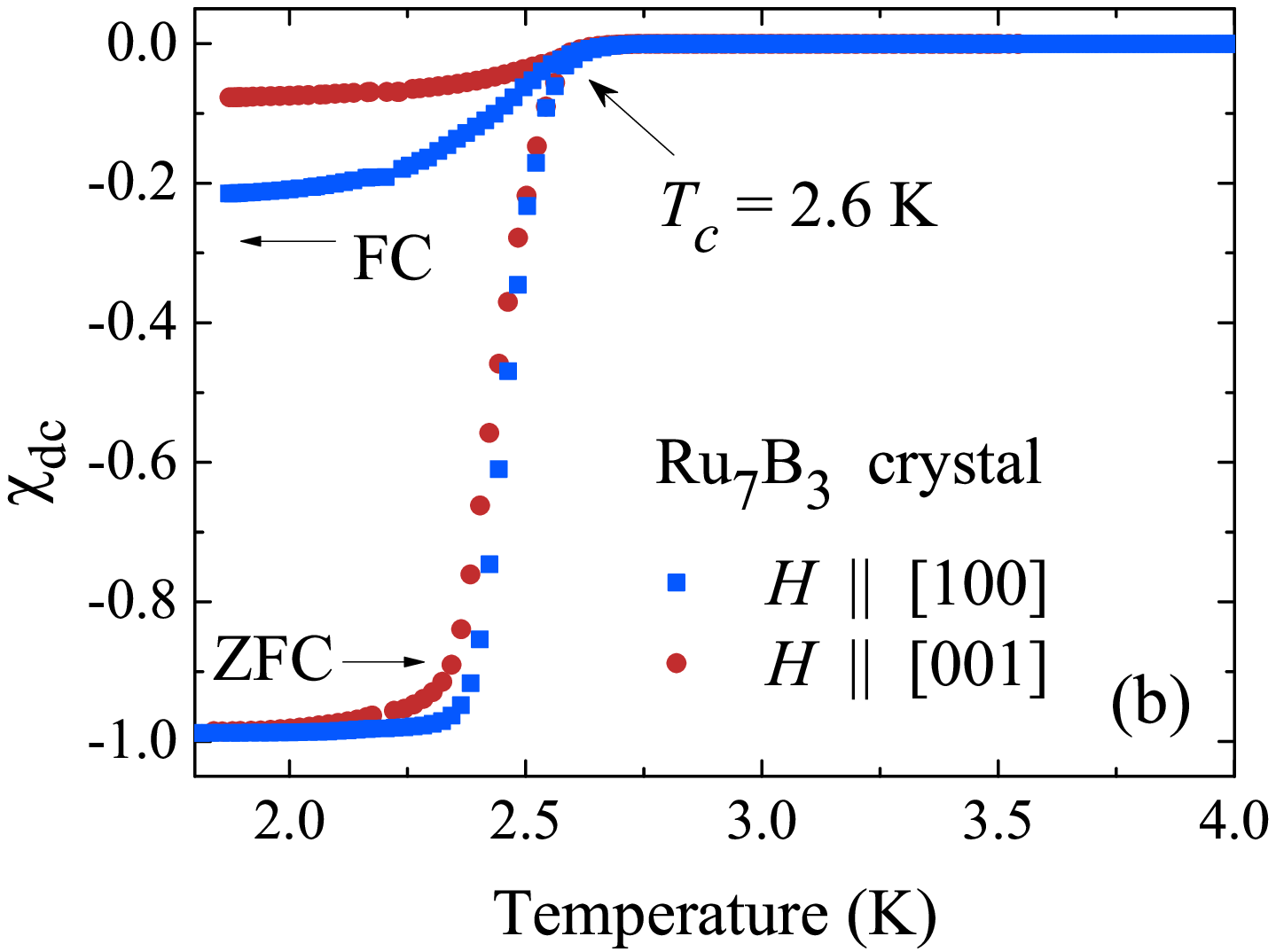}
        \label{fig2b}
    \end{subfigure}
    \caption{Temperature dependence of the magnetic susceptibility ($\chi_\mathrm{dc}$) of (a) polycrystalline {\rub}, and (b) single crystal {\rub}. The magnetic susceptibility was measured in an applied field  of 10 Oe. For the single crystal, the field was applied parallel to the [001] (red circles) or [100] (blue squares)  directions of the crystal . Measurements were carried out in both zero field cooled (ZFC) and field cooled (FC) modes.}
    \label{fig2a2b}
\end{figure}

Figure \ref{fig2a2b} shows the dc magnetic susceptibility as a function of temperature for a crystal of {\rub} as well as for a piece of polycrystalline sample of {\rub}. The onset of the superconducting transition $T_c^\mathrm{onset}$ for the single crystal, determined from the dc magnetic susceptibility measurements, is 2.6~K.  This is different to the $T_c^\mathrm{onset}$ of 3.2~K measured on the polycrystalline ingots used as the starting material for the crystal growth.  Similar differences have also been reported between the $T_c$s of single crystal and polycrystalline material by Kase~\textit{et al}.~\cite{Kase} for crystals grown by the Czochralski technique. In borides, this is often associated with loss of boron during the crystal growth procedure and the presence of boron vacancies in the resultant crystals~\cite{GB}.  In general there is good agreement between the observed transition temperatures and those reported on both polycrystalline samples as well as single crystals grown by the Czochralski method~\cite{Kase}. Magnetisation measurements have been made with a magnetic field applied along two different orientations, [100] and [001].

\begin{figure}[!b]
\begin{centering}
\includegraphics[width=0.7\columnwidth]{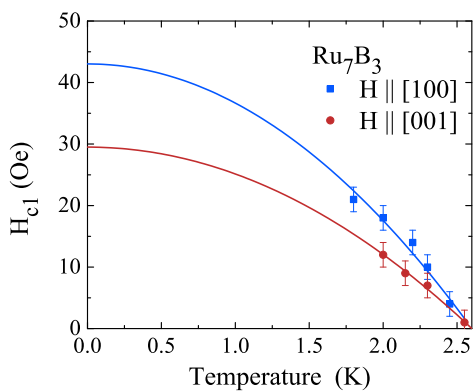}%
\caption{Temperature dependence of the lower critical field $H_{c1}$ for {\rub}. Measurements were taken with the applied magnetic field parallel to the [100] (blue squares) and [001] (red circles) directions of the single crystal. The lines are a fit to the data using  a parabolic $T$ dependence of $H_{c1}$ (described in the text). The best fit values  of $H_{c1}$(0) are 30(2)~Oe for the [100] direction and 43(2)~Oe for the [001] direction.}
\label{fig3}%
\end{centering}
\end{figure}

The lower critical field for single crystal {\rub} was estimated by measuring the magnetisation $(M)$ as a function of applied magnetic field $(H)$ at various temperatures and determining the field at which each $M(H)$ curve deviated from linearity. The $T$ dependence of $H_{c1}$ determined from these measurements is shown in Figure~\ref{fig3} for two different orientations, [100] and [001]. Assuming a simple parabolic $T$ dependence for {$H_{c1}(T)=H_{c1}(0)$~(1-$t^2$)} where $t=T/T_c$, we estimate $H_{c1}(0)$ to be of the order of 30(2)~Oe for the [100] direction and 43(2)~Oe for the [001]. A comparison of our estimates of $H_{c1}(0)$ with the previously reported values for the crystal obtained by Kase~\textit{et al}.~\cite{Kase}, of 68~and~53~Oe respectively, show that our values are slightly lower but the degree of anisotropy observed in both cases is similar ($\sim1.3$~to~1.4).

\begin{figure}[!t]
    \centering
    \begin{subfigure}[h]{\columnwidth}
        \centering
        \includegraphics[width=0.77\textwidth]{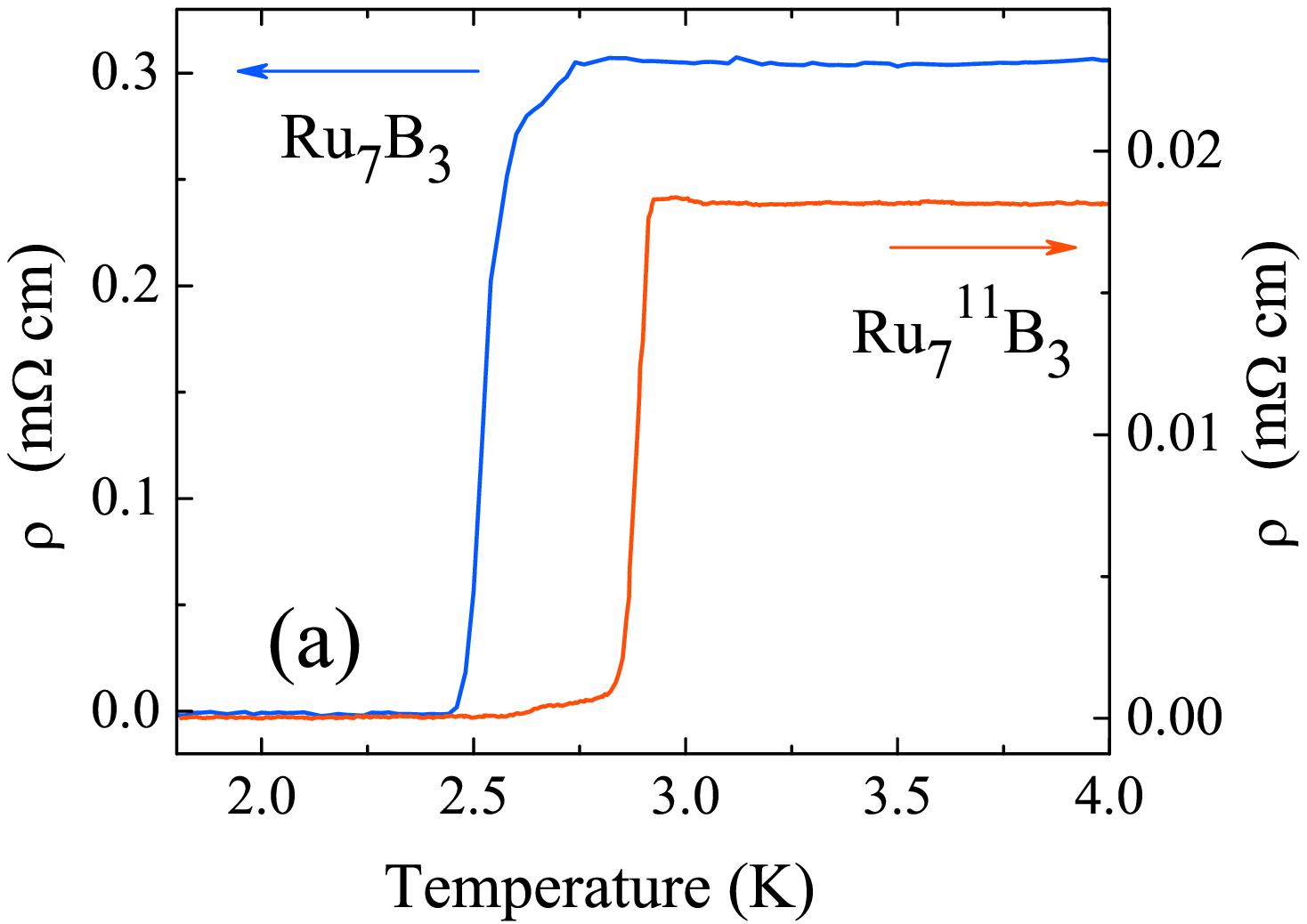}
        \label{fig4a}
    \end{subfigure}
    \begin{subfigure}[h]{0.74\columnwidth}
		\centering
        \includegraphics[width=\textwidth]{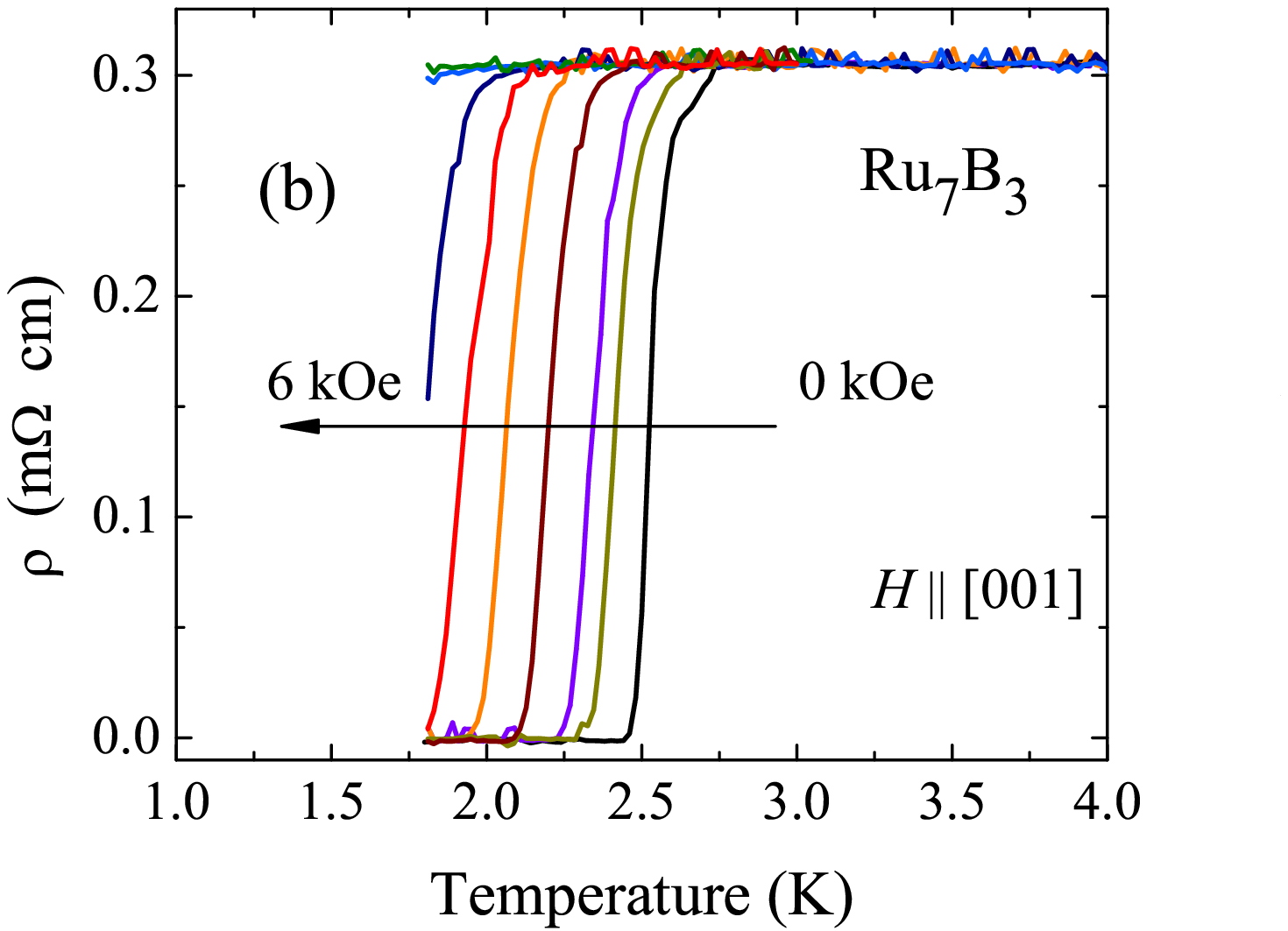}
        \label{fig4b}
    \end{subfigure}
    \caption{(a) Temperature dependence of the resistivity of {\rub}. The {\rub} single crystal shows a superconducting transition  with $T_{c}^\mathrm{zero}=2.5$~K. Also shown is the transition observed for the isotopically enriched crystal, {\eb} (see text). (b)~Temperature dependence of the resistivity of the {\rub} crystal, measured in increasing magnetic fields applied parallel to [100] direction.}
    \label{fig4a4b}
\end{figure}

The superconducting transition measured by resistivity measurements on the crystal of {\rub} is shown in Figure~\ref{fig4a4b}(a).  Also shown is the resistive transition measured on a crystal made starting with the enriched boron isotope, {\eb}. We find that the $T_{c}^\mathrm{onset}$  for the crystal grown using {\eb} isotope is different~(2.8~K) to that observed in the crystal grown using $^{10}$B. Again, this is attributed to the slightly different boron losses encountered during the crystal growth process for the two crystals.

\begin{figure}[!t]
\begin{centering}
\includegraphics[width=0.7\columnwidth]{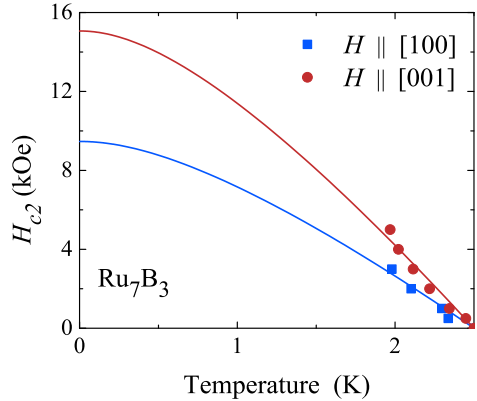}%
\caption{Temperature dependence of the upper critical field $H_{c2}(T)$ of single crystal {\rub}. Measurements were taken with a magnetic field applied parallel to the [100] (blue squares) and [001] (red circles) directions of the single crystal. $T$ was determined from $T_{c}^\mathrm{zero}$. The lines are a fit to the data using a WHH model. The experimental data deviate significantly from this model (details in the text).}
\label{fig5}%
\end{centering}
\end{figure}

The resistive transitions observed in an applied magnetic field are shown in Figure~\ref{fig4a4b}. We observe a gradual reduction in the onset of the superconducting transition as the applied field is increased.  The $T_{c}^{\mathrm{zero}}$ obtained from these curves have been used to determine $H_{c2}(T)$ for the two different directions of the applied field, [100] and [001]. We find that the initial slope of the $H_{c2}$ versus $T$ curve for both the crystal orientations investigated, does not lend itself to fitting with the Werthamer, Helfand and Honenberg (WHH) model~\cite{whh1,whh2}. 
This is similar to the conclusions of Fang~\textit{et al}.~\cite{Fang} for their polycrystalline sample, where fits to both the WHH and the Ginzburg-Landau (GL) models fail, and Kase~\textit{et~al.}~\cite{Kase}, where the WHH model does not fit their data for a single crystal. 
Forcing a fit to our data using the WHH model, we estimate $H_{c2}(0)$ to be 9.5 and 16~kOe for the [100] and the [001] directions, respectively (see Figure~\ref{fig5}).  We observe a greater degree of anisotropy in our $H_{c2}$(0) values than Kase \textit{et al}.~\cite{Kase} who estimated $H_{c2}(0)$ values of 15.8 and 17.2~kOe for the [100] and the [001] directions, respectively, using a GL model. Fang \textit{et al}.~\cite{Fang} report a value of 11~kOe for a polycrystalline sample. Clearly the temperature dependence of the $H_{c2}(T)$ curves requires further detailed analysis. 

\section{Summary and Conclusions}

We have successfully produced large single crystals of {\rub} by the floating zone technique, using an optical furnace equipped with xenon arc lamps. Examination of the crystals by Laue x-ray diffraction indicate that the quality of the crystals is good. Crystals of large volume, free of any contamination can be produced by this technique. The crystals exhibit a superconducting transition at $\sim2.6$~K. Crystals have also been obtained by the same technique using isotopically enriched {\eb} for use in  neutron scattering experiments to probe the vortex state of these superconductors. 

\section{Acknowledgements}

We acknowledge the EPSRC, UK for providing funding (grant number EP/I007210/1). We thank T.E. Orton for technical support and A.D. Hillier (ISIS, STFC, UK) for helpful discussions.  Some of the equipment used in this research at the University of Warwick was obtained through the Science City Advanced Materials: Creating and Characterising Next Generation Advanced Materials Project, with support from Advantage West Midlands (AWM) and part funded by the European Regional Development Fund (ERDF).


\bibliographystyle{model1a-num-names}
\bibliography{ReferencesRu7B3JofCG}











\end{document}